# Diamagnetic Levitation Using High-Temperature Superconducting Wires for Microgravity Research and Mitigation in Human Spaceflight Applications

G. Bruhaug[a*]

[a]University of Rochester, Department of Mechanical Engineering,

235 Hopeman Building, P. O Box 270132,

Rochester, NY 14627 USA

*gbruhaug@ur.rochester.edu

L. Beveridge[b]

[b]LANL Stop E557

PO Box 1663

Los Alamos, NM 87545

beveridge@lanl.gov

A novel use of high-temperature superconducting (HTS) electromagnets for human-sized microgravity research and mitigation is outlined. Recent advances in HTS technology have resulted in electromagnets that potentially could levitate large diamagnetic targets, such as human organs, for additive manufacturing or entire humans for microgravity training. These applications are then used as a springboard to discuss the possibility of active microgravity compensation and "inertial dampeners" for future space travel applications. Finite element simulations are used to check the validity of the designs and motivate future research.

# 1. Introduction

The stable levitation of diamagnetic objects has been studied and demonstrated experimentally but has so far had limited applications. Diamagnetic materials weakly repel any magnetic fields that are applied to them [1], but the repulsion is much weaker than with other magnetic effects such as ferromagnetism. This natural phenomenon can be used to levitate any structure that is primarily diamagnetic, as long as the applied magnetic field is extremely strong.

Until recently, magnets of the needed strength (of the order of several tesla or greater) were either nonexistent or extremely rare. However, the advent of low-temperature superconductors (LTS's) as a commercial technology has made several-tesla magnets quite commonplace. In addition, the rapid development of high-temperature superconductor (HTS) technology has helped push magnetic fields even higher and made the magnets more practical to operate industrially due to the use of more-common coolants.

Diamagnetic levitation was first demonstrated in the 1930s with the levitation of extremely small flakes of graphite [1]. In the early years of the 21st century, more-prominent experiments showed the levitation of a frog and several mice [2]. The animals were unharmed and experienced something akin to microgravity. In addition, the mice were kept in the artificial microgravity environment for extended periods of time to help determine the effects of the magnet and artificial microgravity on their physiology. It is important to note that no adverse effects were seen in the animals placed into the magnets. However, a recent study into the safety of ultrahigh-field MRI (magnetic resonance imaging) (7 T or greater) has brought up possible discomfort concerns [18]. It should be noted, however, that the discomfort felt by some patients is thought to come primarily from the motion through ultrahigh magnetic fields rather than static fields alone. This paper will explore two novel applications of this ultra-high magnetic field technology while also summarizing more near-term applications that could be built with existing technology.

**2. Theory**

*2.1 Diamagnetic Levitation*

Magnetic levitation from repulsion is generally unstable as described by Earnshaw's theorem. It can be shown, however, that with materials that are diamagnetic, levitation is inherently stable [1]. The stability arises from the negative relative permeability of diamagnetic materials. To levitate a diamagnetic object, the following equation must be satisfied:

$$B\frac{dB}{dz} = \mu_0 \rho \frac{g}{x}, \tag{1}$$

where $B$ is the magnetic field, $z$ is the length along the vertical axis in the magnet, $g$ is the gravitational acceleration, $\rho$ is the density of the material, $\mu_0$ is the permeability of free space, and $\chi$ is the magnetic susceptibility of the material. Stability conditions go along with this equation as well to ensure any object will always be "pushed" toward the desired levitation point in the field [1]:

$$\frac{d^2 B^2}{dx^2} > 0, \frac{d^2 B^2}{dy^2} > 0, \frac{d^2 B^2}{dz^2} > 0. \tag{2}$$

The region of stability depends heavily on the design of the magnet but can easily be large enough to trap macroscopic objects, as previously mentioned [1]. A solenoid is not required for this analysis, but for simplicity, our entire analysis assumes a solenoid is used to generate the magnetic field in question. For a reasonably thin solenoid of length $L$, the left-hand side of Eq. (1) can be expressed in a more convenient form as

$$B \frac{dB}{dz} \approx B^2/L. \tag{3}$$

As a result, we can find a simple expression for the magnetic field required to induce or cancel an acceleration of $n*g$, where $n$ is a multiple of the gravitational acceleration $g$:

$$B = \sqrt{L \mu_0 \rho \frac{ng}{\chi}}. \tag{4}$$

Levitation will occur at the inflection point in the center of the solenoid [1]. To be levitated at 1 $g$, water and carbon have $B^2/L$ values of roughly 1400 T$^2$/m and 375 T$^2$/m, respectively. All of this assumes the acceleration (or gravity) is parallel to the solenoid axis ($z$ axis in these equations). A simple diagram of the levitation configuration can be seen in Fig. 1.

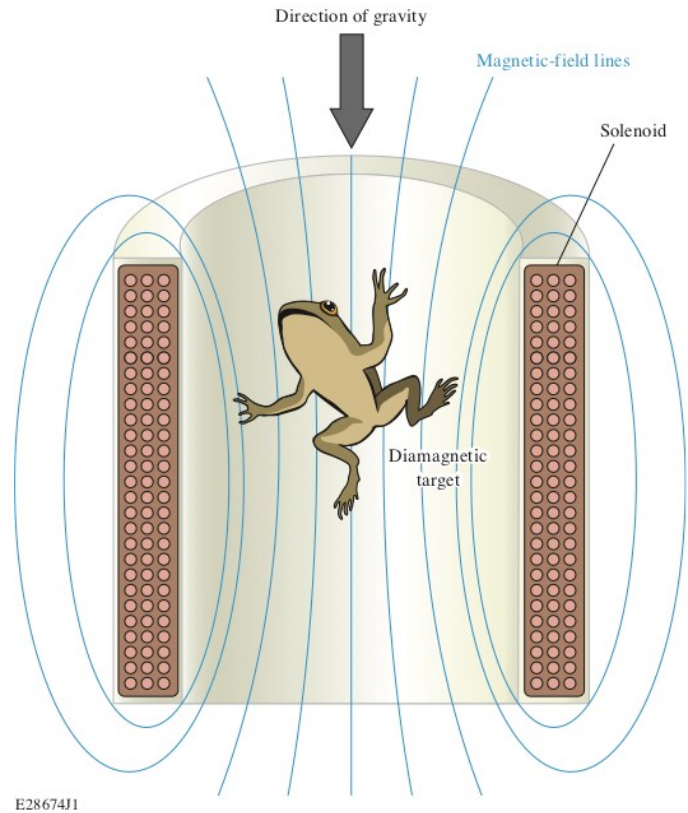

*Figure 1. Diagram of simple diamagnetic levitation configuration.*

## 3. Applications

*3.1 Microgravity Simulation for Biological Applications*

One potential application of diamagnetic levitation is the production of complex biological systems that prove difficult to manufacture in the full gravitational pull of Earth. An application that has already been explored is growing protein crystals [3]. These experiments were able to replicate microgravity within a small volume, allowing crystals to grow without the effects of gravity. These experiments were also able to effectively invert the force on the crystals, causing them to grow as if gravity were reversed in direction. The ability to grow biological samples or implants as if they were in microgravity without requiring an expensive launch to orbit could be a tremendous opportunity for medicine and biotechnology.

Another unexplored field could be the 3-D printing of replacement organs. Organs like the human heart are extremely difficult to print in 3-D in a gravitational field [10], and experiments have already taken place to determine the usefulness of low-gravity manufacturing [10]. Rather than launch costly equipment and subject new hearts to the stress of landing and recovery, we propose building them inside of an extremely strong electromagnet. An average adult male heart masses 331 g [11] and has a very consistent composition. Therefore, a heart could be levitated with a far smaller magnet (possibly 20 T or less) than will be discussed later. These magnets may soon be easily built with contemporary HTS technology [13]. The primary difficulty in this method will be designing organic 3-D printing equipment that can function inside such a large magnetic field.

*3.2 Microgravity Simulation Facility*

A more exotic application of diamagnetic levitation is the on-Earth simulation of microgravity for people. Currently, microgravity simulation facilities consist of either neutral buoyancy training in swimming pools or small (20-s to 30-s) excursions during a parabolic flight [6]. The major drawback of the neutral buoyancy technique is the inability to simulate the day-to-day life challenges of a microgravity environment, such as eating and hygiene; whereas the parabolic flight method is not able to provide an adequate simulation for any task that lasts longer than half a minute. In addition, the parabolic flight method is used for some small microgravity research experiments but suffers from the short duration of each excursion.

Using diamagnetic levitation to create a microgravity facility would provide several benefits over the contemporary methods of microgravity simulation. First, the simulation could last as long as needed to provide the required training or research data. Any magnet used for this technique would be superconducting and, as such, would only need to be continuously cooled after the necessary currents are applied. This technique is commonly done with MRI machines [7] and is analogous to charging a battery. Second, the simulation facility could easily be made to accommodate microgravity hygiene, sleeping, and dining facilities. This would allow future astronauts to be completely familiar with the myriad array of challenges that one encounters in microgravity before ever stepping foot into a rocket. The final advantage is the ability to alter the simulated gravity from microgravity to full Earth gravity and conceivably even to higher accelerations if needed. By applying less current to the diamagnetic levitation facility, the gravity simulation could be made to mimic the conditions on the Moon or Mars (both future targets for human exploration). In addition, the orientation of the field could be reversed and used to push on the trainees to increase the perceived gravity. This has implications that will be discussed later.

The approximate size of the facility will be discussed in the next section; however, any permanent ground installation has far more freedom in facility size, mass, and power consumption than any space flight hardware. It would be advantageous to build a larger, cheaper facility on the ground in the near term to help determine potential problems for future flight hardware.

*3.3 Gravity Simulator*

Another possible application would be the generation of a gravity-like force on a spacecraft or space station. Simulating Earth-like gravity would minimize the adverse effects of microgravity on the body [4,5,8]. It is likely not practical to simulate gravity in the entire spacecraft due to weight and volume limitations. However, the reverse of the microgravity apparatus described in the previous section to train astronauts could be put onboard a spacecraft so that astronauts could sleep or exercise in simulated gravity for part of their day. Although the magnets would be quite heavy, they have the potential to be lighter than a rotating habitat due to the minimum size constraint placed on a rotating habitat to minimize crew discomfort [4,5]. Furthermore, other alternatives such as vacuum treadmills do not act on the entire body and, consequently, do not fully mitigate the physiological effects of microgravity.

*3.3.1 Magnet Size*

The magnetic-field requirements for both the microgravity facility and the gravity simulator are the same, assuming the same internal volume. Using Eq. (4), with $n = 1$ we can find how strong the solenoid must be for a given length (plotted in Fig. 2).

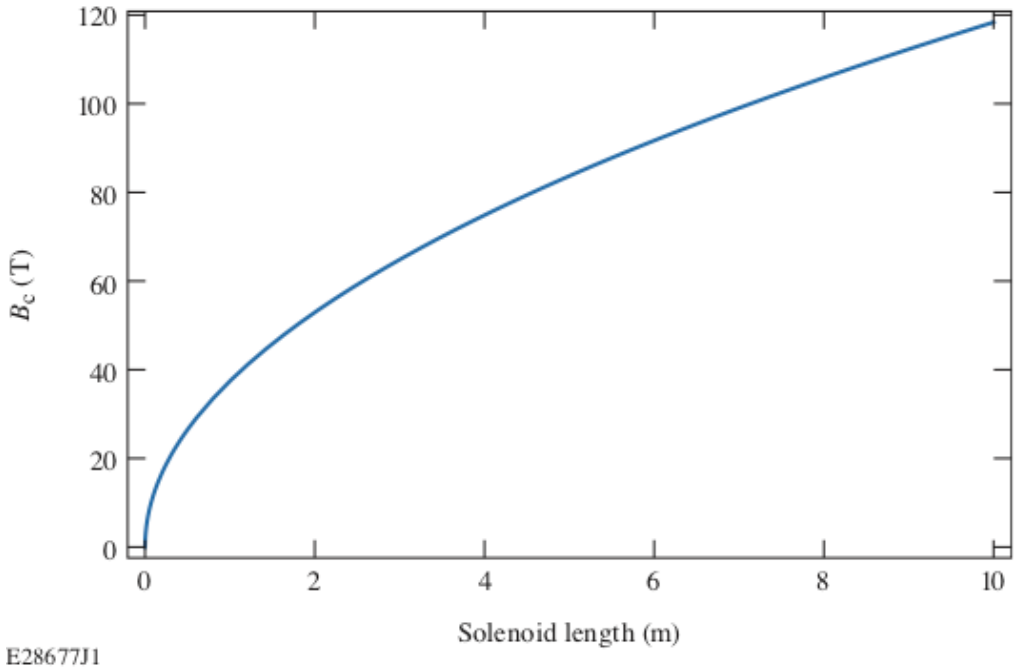

*Figure 2. Plot of magnetic-field strength to produce effective acceleration of 1 g versus solenoid length.*

Although the solenoid must be longer than the useful simulation volume, it would still be large enough to hold one to two people with magnetic-field strengths achievable today. The following simple example will help illustrate the size and current requirements for a roughly 1-g gravity simulator.

To determine the field strength, we will assume that the internal diameter will be 2 m and the internal height will be 2.5 m, with the extra height chosen to help avoid field fringe effects and have room for any needed internal equipment (such as for exercise or research). This provides the astronaut(s) with 7.85 m$^3$ of internal volume and 3.14 m$^2$ of useable floor. To provide the approximately 1 g of force needed, the solenoid will need a centerline B field of roughly 50 T.

Current-day commercial ReBCO HTS cables are rated to a maximum current density of 2 MA/cm² [12,13] when cooled to 4.2 K with a perpendicular magnetic field of 50 T. The commercially available cables have a ReBCO layer that is 1.6 μm thick and 12 mm wide (Fig. 3). This translates to a critical current of 384 A (amperes) in the cables. To add a factor of safety, the maximum current allowed will be assumed to be 90% of the critical current, which is 345.6 A.

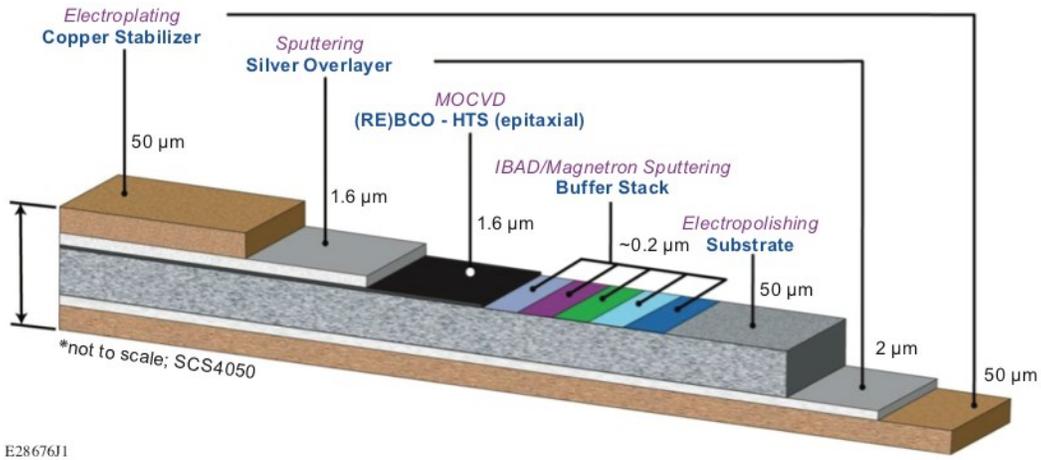

*Figure 3. Cross section of commercial ReBCO HTS cable. [12].*

A thin-walled, finite solenoid approximation will be used to determine the total number of turns the solenoid requires [Eq. (5)]:

$$B = \frac{\mu_0 I N}{\sqrt{L^2 + 4r^2}}, \qquad (5)$$

where $\mu_0$ is the magnetic permeability of free space, $I$ is the current in the solenoid, $N$ is the number of turns of the wire, $L$ is the length of the solenoid, $r$ is the inner solenoid radius, and $B$ is the magnetic-field strength. The equation can then be rearranged to solve for $N$ using B = 50 T, $I$ = 345.6 A, L = 2.5 m, and $r$ = 1 m. This yields a result of 368,500 turns of the cable. The commercial ReBCO cable is 155.4 $\mu$m thick and 12 mm wide. Therefore, it will have to be made into a multilayer solenoid to reach the required number of turns. The multilayering begins to stretch the thin-walled assumption; however, it will suffice for this example.

The ReBCO cable will be turned into 23 layers of cable around the central cylindrical region. This will result in a magnet that is 2.5 m tall and 2.552 m in outer diameter. The magnet will need over 2000 km of ReBCO cable. In addition, if the cables are assumed to have roughly the density of copper, which seems to be a safe assumption since the bulk of the cable is solid copper [12,13], the total cable mass is found to be approximately 102 tons.

Although these masses are too large for practical use in aerospace applications at present, there are several things to consider that could improve the situation. First, HTS wire development is still in its infancy and new developments could occur. In addition, the superconductor composes only 1% of the volume of the commercial cables [12]. A large amount of mass could be saved by cutting down the total volume of copper and silver-backing material in the ReBCO cables. A second consideration is the relationship that critical current has with operating temperature and perpendicular magnetic field. If the magnetic field was lowered to allow for lower gravity levels, or the wires cooled below 4.2 K, the critical current could be increased and the total wire mass lowered. A final consideration is the development of new HTS materials, which are being discovered at a prodigious rate [19]. It is not presumptuous to assume that a better material could be developed to help increase critical currents and lower total magnet size. It is also worth noting that the non-space travel applications suffer from none of the mass concerns and could be built today.

*3.4 Inertial Dampener*

The ability to simulate or effectively eliminate the perception of gravity implies that it might also be used to dampen acceleration on a person in a rapidly accelerating vehicle. Because the magnetic field acts on all tissues to essentially the same degree (because they are mostly water), an effective acceleration could be made to exactly counter the acceleration of a vehicle; consequently, the occupant would not feel the effects of the acceleration. This machine will still conserve momentum by applying the negated force to the magnet itself and then to the surround support structure. Such a machine would provide huge advantages over more traditional acceleration mitigation methods in crew safety. There would be no risk of losing consciousness or vision during the most extreme maneuvers and cardiac threats would be alleviated. Using Eq. (4) again, we can plot the magnetic field that would be required from a 3-m-long solenoid for a range of acceleration-induced forces (shown in Fig. 4).

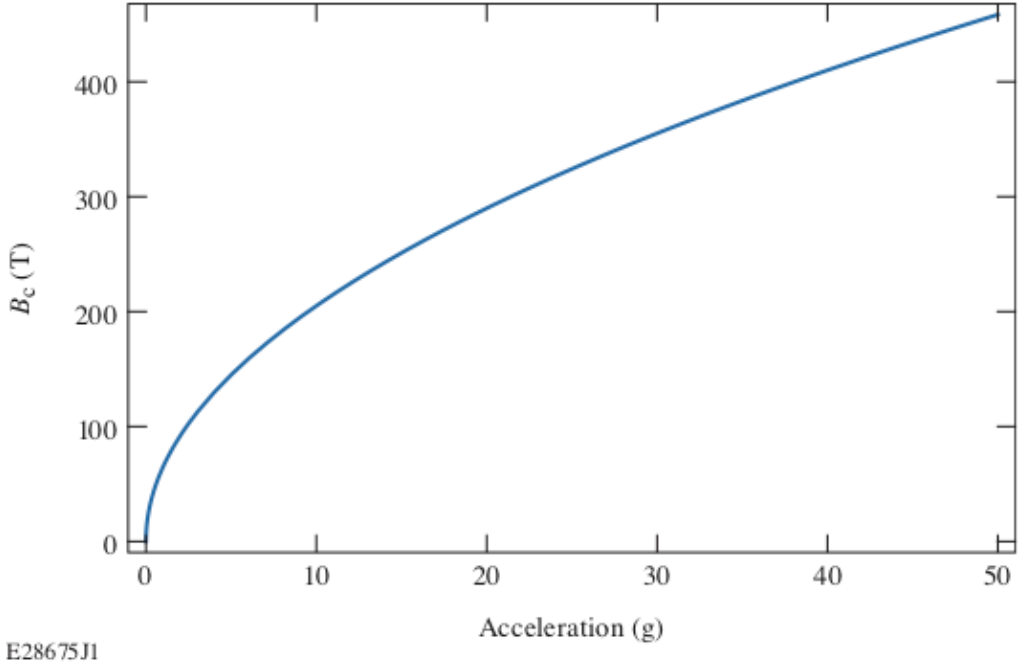

*Figure 4. Plot of magnetic-field strength to offset vehicle acceleration felt by occupant.*

Clearly, this requires enormous fields, but even these may be possible with future superconductors [15]. Of course, since the field is exerting such a large effective acceleration, differences in the magnetic susceptibility between different types of tissue could produce significant forces between them. Bone and soft tissue likely have the greatest difference in susceptibility due to the solid composition of the bone. The susceptibility for soft tissue is still assumed to be the same as water, but bone is much different, with a magnetic susceptibility of about $-1.1 \times 10^{-6}$ (SI units) and a density of 1500 kg/m³ [9], whereas water has a magnetic susceptibility of $-8.8 \times 10^{-6}$ and 1000 kg/m³. Using Eq. (4), we can solve for the difference in acceleration between them (again, $n$ is multiples of $g$):

$$|a_{\text{water}} - a_{\text{bone}}| = \left| n \left( 1 - \frac{\rho_{\text{water}} \chi_{\text{bone}}}{\rho_{\text{bone}} \chi_{\text{water}}} \right) \right|, \qquad (6)$$

which is plotted in Fig. 5.

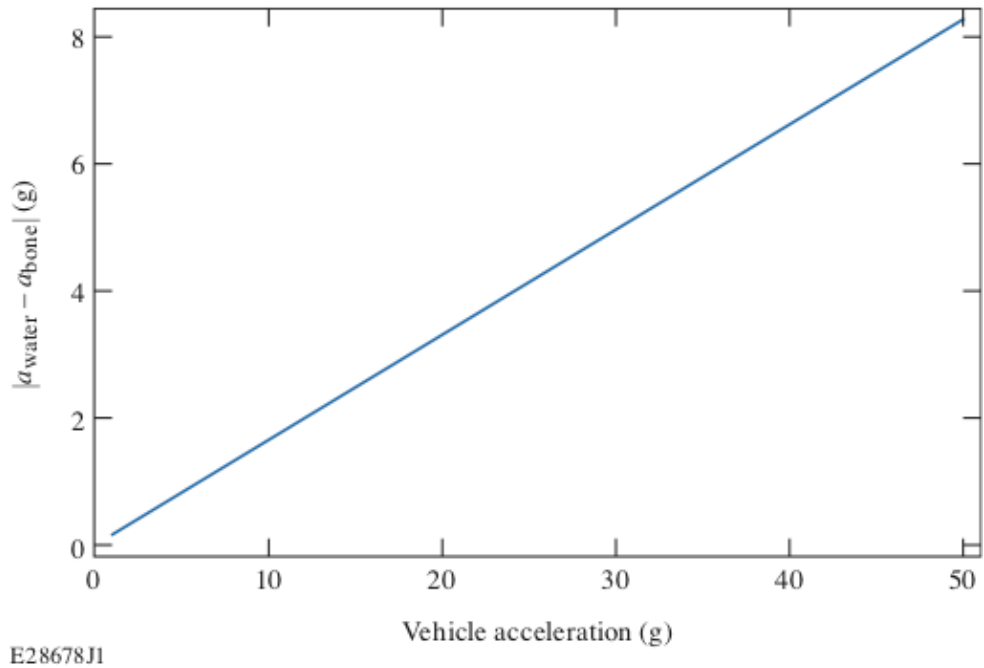

*Figure 5. Plot of effective acceleration difference between soft tissue and bone.*

This indicates that even under ideal conditions, with an ideal magnetic arrangement, the maximum acceleration that can be mitigated for a human is likely not much more than about 50 *g*'s. As of yet, there is no obvious application for such an "inertial damper." If something akin to a mass-driver or other ballistic space-launch system is developed [20], however, this technology may make it feasible to launch humans using a shorter mass driver than normally expected. Additionally, this technology could prove beneficial for future spacecraft that use advanced high-thrust propulsion. For instance, a magnetic inertial dampener would help alleviate the extreme oscillatory accelerations that the occupants would experience during the engine firing of a nuclear pulse–propelled vehicle [14].

## 4. Simulations

*4.1 Finite Element Simulations*

As a check on our calculations, we reproduced the levitating frog experiment by simulating it using an open-source finite element analysis software package known as *Finite Element Magnetic Methods* (FEMM) [17]. An example of the axisymmetric FEMM model and output can be seen in Fig. 6. The small half circle in the center of the simulation represents the frog. In the simulation the frog is treated as having the magnetic permeability of water as was done by Simon and Geim [1]. The off-center rectangles represent an axisymmetric slice of the two-part Bitter solenoid used, and the lines emanating from the center of the picture represent the magnetic-field lines. The color gradient shows magnetic field strength in Teslas.

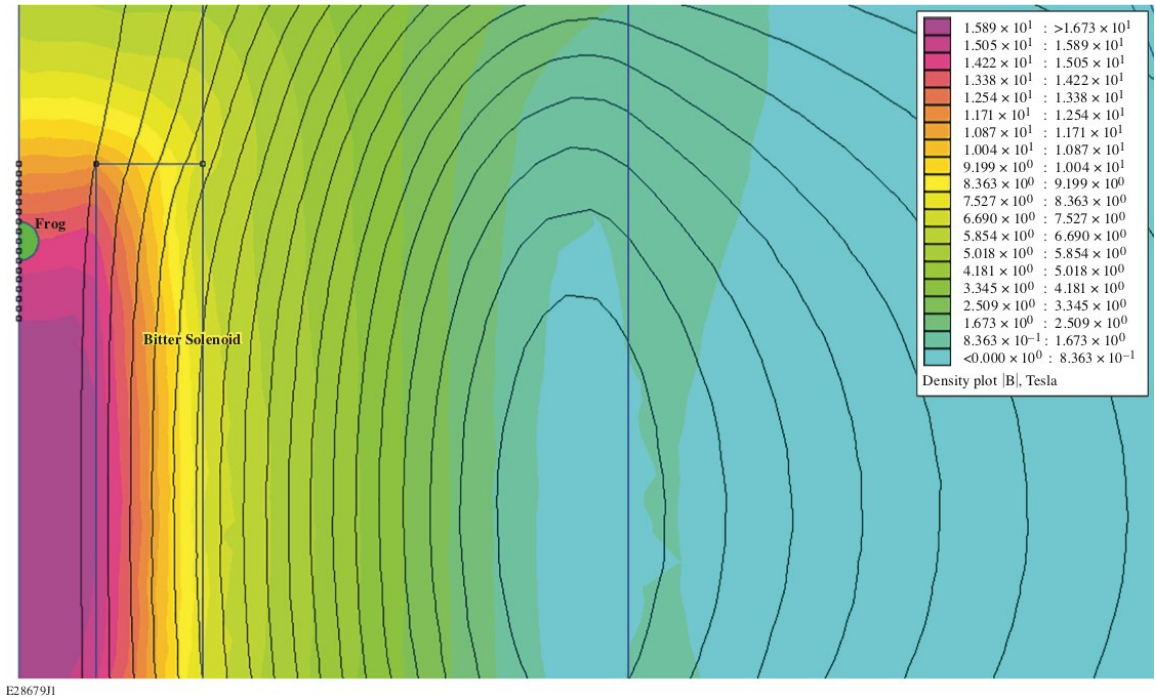

*Figure 6. Finite element simulation of the diamagnetic levitation of a frog.*

The Bitter solenoid was simulated with an equivalent number of turns of wire and the same current as reported in the experiment. The center-line magnetic field was found to be roughly the 16 T reported by Simon and Geim [1]. The force on the sphere of water was then calculated via a function in FEMM and checked against the force of Earth's gravity on an equivalent sphere of water. The position of the sphere was then adjusted until the upward force was found to equal the downward force of gravity. The downward force of gravity is negated at a position roughly 70 mm from the center of the solenoid, which agrees with the 69±1 mm reported [1].

A similar simulation was performed for the gravity simulator as described above. A 0.1-m in radius, 1.6-m-long rod of water (to approximate a human) was placed in the center of a large solenoid, which can be seen in Fig. 7. The half-circle lines on the edge of the solenoid represent the Dirchlet boundaries of the simulation.

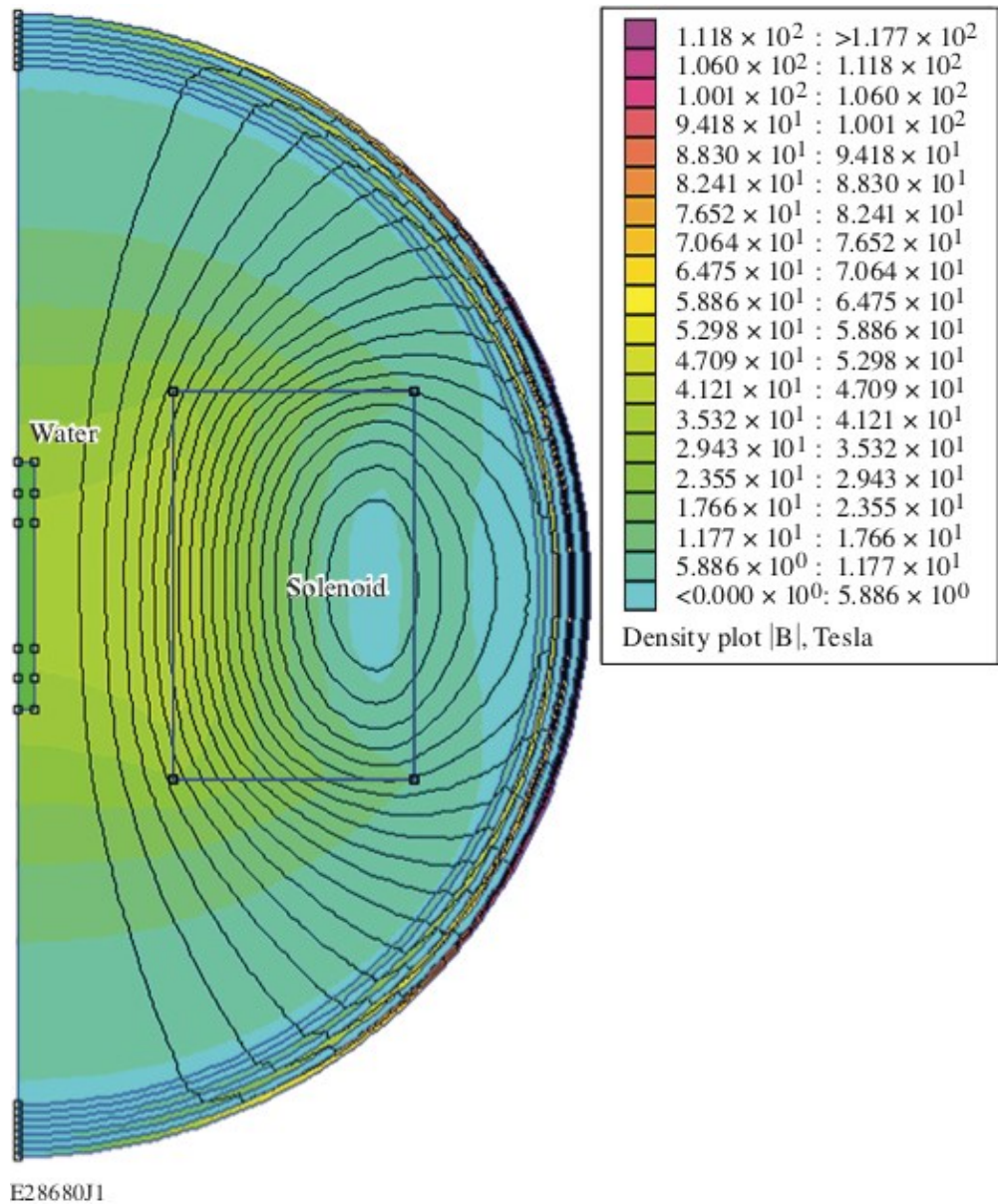

*Figure 7. Gravity simulator with a water rod human stand-in.*

The water rod would mass 50 kg, yet the FEMM simulation shows that the magnet would provide over 6× more force than the required ~500 N needed to offset gravity. This result shows that a smaller magnet, or one using less current, could be used to balance a human test subject against the force of gravity. Additionally, the predicted magnetic field on target is much higher than a simple back-of-the-envelope calculation would lead one to think. Further optimization may be had by noting the unique position of the levitating "frog" in Fig. 6. A magnet with a sharp gradient in its construction may achieve similar levitation results with far less material and lower magnetic fields. Preliminary simulations have shown that the magnet mass by a factor of 10 and the required field by a factor of 25. More study will be needed to create optimal magnets and prevent force gradients from forming across the test subject.

## 5. Conclusions

Modern HTS technology allows one to construct extremely large and high-field magnets. These magnets have many conventional applications; however, several unconventional applications have not been fully explored. Large magnets generating 20 T plus fields could be used for advanced biotechnology manufacturing, up to and including the 3-D printing of human hearts. Even larger magnets could be used to simulate variable gravity fields here on Earth for training future astronauts. This technology exists today and could be built with commercial technology. If these magnets can be reduced in mass, they could be used on future spacecraft to simulate Earth gravity to help alleviate the health problems associated with low gravity. Simple simulations show that there seems to be a viable path forward to reduce the required magnet masses and currents. Even more-advanced magnets could be used as inertial dampeners to help human occupants withstand the acceleration of advanced propulsion systems, such as nuclear-pulse propulsion and mass drivers.

**Figures**

Figure 1. Diagram of simple diamagnetic levitation configuration.

Figure 2. Plot of magnetic-field strength to produce effective acceleration of 1 g versus solenoid length.

Figure 3. Cross section of commercial ReBCO HTS cable. [12].

Figure 4. Plot of magnetic-field strength to offset vehicle acceleration felt by occupant.

Figure 5. Plot of effective acceleration difference between soft tissue and bone.

Figure 6. Finite element simulation of the diamagnetic levitation of a frog.

Figure 7. Gravity simulator with a water rod human stand-in.